\documentclass[
amsmath,amssymb,
aps,
prl,
twocolumn, superscriptaddress, showpacs
]{revtex4-1}

\usepackage{graphicx}% Include figure files
\usepackage{dcolumn}% Align table columns on decimal point
\usepackage{bm}
\usepackage{mathtools}
\usepackage{xcolor}
\usepackage{color}
\usepackage[braket, qm]{qcircuit}
\graphicspath{{pics/}}

\def \bra#1{\mathinner{\langle{#1|}}}
\def \ket#1{\mathinner{|{#1}\rangle}}

\def \red #1 {\textcolor{red}{#1}}

\usepackage{tikz}
\usepackage{circuitikz}
\usetikzlibrary{backgrounds,fit,decorations.pathreplacing,calc,backgrounds}  % TikZ libraries

\begin{document}

\title{Pure State Tomography with Fourier Transformation}% Force line breaks with \\

\author{Yu Wang}
\affiliation{Yanqi Lake Institute of Mathematical Sciences and Applications, Beijing, 100407, China}

\author{Keren Li}
\email{likr@pcl.ac.cn}
\affiliation{Peng Cheng Laboratory, Shenzhen 518055, China}

\date{\today}
\begin{abstract}

Extracting information from quantum devices has long been a crucial problem in the field of quantum mechanics. By performing elaborate measurements, quantum state tomography, an important and fundamental tool in quantum science and technology, can be used to determine unknown quantum states completely.
In this study, we explore methods to determine multi-qubit pure quantum states uniquely and directly. 
Two adaptive protocols are proposed, with their respective quantum circuits. 
Herein, two or three observables are sufficient, while the number of measurement outcomes are either the same as or fewer than those in existing methods.
Additionally, experiments on the IBM 5-qubit quantum computer, as well as numerical investigations, demonstrate the feasibility of the proposed protocols. 

\end{abstract}
\maketitle

\emph{Introduction}
--The superposition and entanglement properties of quantum states account for the various elegant quantum algorithms \cite{Deutsch1985,Deutsch1992,Shor1994,Grover1997,Chuang1998} and  efficient communication protocols in quantum information processing \cite{Ekert1991,Bennett1992,Bennett1993,Gisin2002,Gisin2007}. 
A quantum bit resembles an arbitrary state of a spinning coin. 
A qubit (pure state) is defined as $a|0\rangle+b|1\rangle$, where $a$ and $b$ are complex numbers, and $|a|^2+|b|^2=1$. 
When a quantum computer with an initial state $|0\rangle^{\otimes N}$ is operated,  we obtain an $N$-qubit pure quantum state, if unaffected by noise. 
All of its possible states can be defined with a superposed mathematical form, given by $\sum_{k=0}^{2^N-1}a_k|k\rangle$, where $\{a_k\}$ are complex numbers. 
The readout of the coefficients $\{a_k\}$, is important in several applications. This function is a benchmark in verifying the performance of quantum machines, producing desirable quantum states \cite{Eisert2020}. For example, fidelity is calculated to confirm whether the state currently being transported is close to the initial state, for certain quantum communication protocols \cite{Riebe2007,Baur2012}. Moreover, the entropy and the entanglement measure of the states can be directly computed, provided the states are known \cite{Peres2002,Islam2015,Torlai2018,Steffen2006}.

Quantum state tomography aims at obtaining an unknown quantum state by measuring a group of identical quantum states \cite{Vogel1989,Leonhardt1995,White1999,James2005}. First, a few applicable observables are designed and implemented (or physical quantity measurement settings are designed) to measure the quantum states. 
Then, the frequency of outcomes in each measurement is recorded to compute the probability distributions of measurements, or the expectation values of the observables. 
With these data, different reconstruction methods can be used to estimate the unknown quantum states. Certain important indicators relevant to the time and cost of the process exist, such as the number of observables, complexity of the reconstruction algorithms, and number of measurement outcomes.

Pursuing a minimal number of observables has a long history. 
With the inception of quantum mechanics to define quantum states, the problem of unique determination has been generated. Pauli examined whether a wave function can be determined using position and momentum distributions \cite{Pauli1933allgemeinen}. However, this is impossible as identical probability distributions may correspond to different wave functions \cite{Trebino2000}. Furthermore, Asher Peres considered a finite version, as described in his book \cite{peres2006quantum}. A wave function is analogous to a pure $d$-dimensional quantum state (qudit). The position and momentum observables are analogous to the two orthonormal bases $\mathcal{B}_0$ and $\mathcal{B}_1$, which can be transformed by performing Fourier transformation. Peres conjectured that the two orthogonal bases mentioned above would suffice, except in the case of an ambiguous set. Flammia et al. proved that an ambiguous set cannot be a measure zero set \cite{flammia2005minimal}. According to some of the previous studies \cite{Vogt1978,Moroz1983,Moroz1994,Corbett2006,Heinosaari2013,Jaming2014uniqueness,Carmeli2016}, four orthonormal bases are sufficient to uniquely determine an unknown pure qudit. Although the complex mathematical expressions of the projected basis states \cite{Jaming2014uniqueness,Carmeli2016} hinder the physical implementation, a constant number of observables to uniquely determine the pure quantum states is still attractive, especially in an exponential dimension $d=2^N$, in a qubit system.
Decreasing the number of observables is beneficial in physical implementation. The changes in different physical observables are caused by the changes in the measurement setups, which could result in unwanted alterations to the data, prolonged measurements, and impaired noise assumptions \cite{Oren2017}. 
Therefore, it is preferable to carry out the physical implementation with as few observables as possible.

The complexity of reconstruction algorithm affects runtime to process estimation of outputs of certain quantum device, for example, the 10-qubit experiment in a superconducting platform~\cite{Song2017}. Instead of performing a complex computation using the entire data to determine the perfectly matched one, a novel concept of direct reconstruction is proposed herein. Sequential weak and strong measurements are enhanced to directly determine the value of each density matrix for the mixed quantum states~\cite{Lundeen2011, Lundeen2012,Thekkadath2016,Calderaro2018,Zhang2019}.

The number of measurement outcomes is important as it affects the time consumed to obtain all experimental data. Although using as few measurement outcomes as possible is preferable, a lower limit exists, because the number of outcomes strongly influences the ability of the protocol to determine a pure qudit. It has been proved that there should be at least $4d-5$ ($5d-7$) outcomes to uniquely determine a pure qudit among  the pure (mixed) quantum states \cite{Heinosaari2013,Chen2013}. With the adaptive strategy, a measure zero set from all the pure states can be neglected, and thus, only $3d-2$ outcomes are required \cite{flammia2005minimal, Finkelstein2004,wang2018pure}. 
When a qudit is measured using an observable, $d$ outcomes may appear if there are no ancilla qubits, with each outcome corresponding to a projected eigenstate. The different eigenstates of an observable should be orthogonal. This constraint makes the observable design more complicated. 
Goyeneche et al. designed five adaptive orthogonal bases corresponding to $5d$ eigenstates, to uniquely determine an unknown pure qudit state \cite{PhysRevLett.115.090401}. Zambrano et al. constructed three orthogonal bases, where $2d$ eigenstates were used to determine $2^{d-1}$ candidates and the rest of the projected eigenstates were used to find out an estimation through a likelihood function \cite{Zambrano2020}. Accordingly, the following question needs to be addressed: Is it possible to use a few precise measurements to uniquely and directly determine an unknown pure quantum state with fewer number of observables?

In this study, we added an auxiliary qubit and further employed the adaptive strategy to address this challenge. Herein, two protocols are proposed to uniquely and directly determine finite-dimensional pure quantum states with two or three observables. 
The measurements projected onto the orthogonal basis $\mathcal{B}_0$ are used to determine the amplitudes. 
One or two observables connected via a partial Fourier transform are constructed to determine the phases, which can additionally overcome the frequent changes in the measurement setups. 
Besides, the number of measurement outcomes for protocol 2 is decreased to $80\%$ of the outcomes for the protocols proposed by Goyeneche et al. 
Moreover, we present the circuit implementation of $N$-qubit pure states. Two Fourier transformations between the auxiliary qubit and the $N$-th qubit are required, in addition to a global shift operation. 
Furthermore, details of the numerical experiments on mixed quantum states and the experimental demonstration on IBM Quantum Experience are provided, thereby proving the feasibility of both the proposed protocols.

\emph{Main result}
--An $N$-qubit pure state $|\phi\rangle$ in $d=2^N$-dimensional Hilbert space $\mathcal{H}_d$ is given by,
\begin{equation}
 |\phi\rangle=\sum\nolimits_{k=0}^{d-1}a_{k}e^{i\theta_{k}}|k\rangle.
 \label{keyform}
\end{equation}
where $a_k$ and $\theta_k$ are the \emph{amplitude} and \emph{phase} of $|\phi\rangle$, respectively, with $a_{k}\ge0$ and $\theta_{k}\in[-\pi,\pi)$.  

The canonical basis $\mathcal{B}_0=\{|0\rangle,\cdots,|d-1\rangle\}$ is sufficient to determine the amplitudes, $\{a_k\}$. 
Consider that the measurements are carried out repeatedly $M$ times. 
 $n_k$ is recorded as the frequency of the state, collapsed into $|k\rangle$. 
Based on the Born rule, when $M$ is sufficiently large, probability $P_k=|\langle k|\phi\rangle|^2=n_k/M\approx a_k^2$.

The Fourier transformation circuit of $\mathcal{B}_0$ is direct and has many applications such as Shor's algorithm \cite{Shor1994} and HHL algorithm \cite{Harrow2009}. 
But it cannot uniquely and directly determine the phases $\theta_k$ \cite{flammia2005minimal}. 

Therefore, we introduce an auxiliary qubit. 
A new orthogonal basis $\mathcal{C}_1$, or the two orthogonal bases $\mathcal{D}_1$ and $\mathcal{D}_2$ can uniquely and directly determine the phases. The coding rule is presented below.  
The basis states of the compound system $\mathcal{H}_2\otimes\mathcal{H}_d$ are $\{|0\rangle|k\rangle,|1\rangle|k\rangle:k=0\cdots,d-1\}$. 
The states $|j\rangle|k\rangle$ are encoded as $|jd+k\rangle$ of $\mathcal{H}_{2d}$, where $j\in\{0,1\}$, $k\in\{0,\cdots,d-1\}$.

\begin{figure}[ht]

\centering
\includegraphics[scale=0.6]{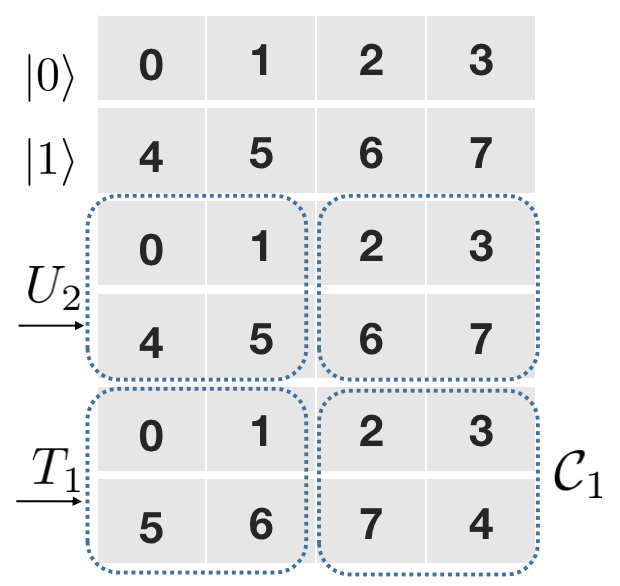}
\caption{Construction of the basis $\mathcal{C}_1$. 
This picture shows the four-dimensional case, i.e., $d=4$, given in Eq.(\ref{c1}).
The basis states of compound system $\mathcal{H}_2\otimes{\mathcal{H}_4}$ are $\{|0\rangle,|1\rangle,\cdots,|7\rangle\}$.
Firstly, a partial Fourier transformation on states $\{0,1,4,5\}$ and $\{2,3,6,7\}$ is performed. 
Then, a conditional shift operation $T_1$ is carried out on the evoluted states, $T_1=|0\rangle\langle 0|\otimes I +|1\rangle\langle1|\otimes\sum_{k=d}^{2d-1}|d+(k\oplus 1)\rangle\langle k|$. 
}
\label{pic-c1}
\end{figure}

\textbf{Protocol 1}
--A new projective measurement onto the orthonormal basis $\mathcal{C}_1$ is sufficient to determine the phases uniquely and directly. 
\begin{eqnarray}
\mathcal{C}_1&=&\{|k,1\rangle_f,|k,2\rangle_f,|k,3\rangle_f,|k,4\rangle_f:0\le  k\le \lfloor\frac{d-2}{2}\rfloor\}.
\label{c1}
\end{eqnarray}
%To obtain these bases states, we rearrange states  $\{|0\rangle,|1\rangle,\cdots,|2d-1\rangle\}$ into several  4-elements set and then operate  Fourier transformation at each set. 
Fig.(\ref{pic-c1}) illustrates the transformation from $\{|0\rangle,|1\rangle\}\otimes \mathcal{B}_0$ into $\mathcal{C}_1$.
Generally, $|k,1\rangle=|2k\rangle$, $|k,2\rangle=|2k+1\rangle$, $|k,3\rangle=|2k+ 1+d\rangle$, and $|k,4\rangle=|2k\oplus 2+d\rangle$. The symbol $\oplus$ represents the modulo $d$ operation. 
The subscript $_f$ is used to denote the FT. The states can be represented as follows:

\begin{eqnarray}
|k,1\rangle_f             &   =  &  (|k,1\rangle+|k,2\rangle+|k,3\rangle+|k,4\rangle)/2,     \nonumber \\
  |k,2\rangle_f           &   =  &  (|k,1\rangle+i|k,2\rangle-|k,3\rangle-i|k,4\rangle)/2,   \nonumber  \\
  |k,3\rangle_f   &   =  &  (|k,1\rangle-|k,2\rangle+|k,3\rangle-|k,4\rangle)/2,  \nonumber  \\
  |k,4\rangle_f  &   =  &  (|k,1\rangle-i|k,2\rangle-|k,3\rangle+i|k,4\rangle)/2. \nonumber \\
  \label{cc1}
\end{eqnarray} 

For the reconstruction process, we consider $|\Phi_0\rangle=|0\rangle|\phi\rangle$ and $|\Phi_1\rangle=|1\rangle|\phi\rangle$. 
 Repeatedly measure the states $|\Phi_0\rangle$ and $|\Phi_1\rangle$ with the projective measurement on $\mathcal{C}_1$ and record the frequency of measurement outcomes. 
The probability is denoted as $P_{|k\rangle_f}$ ($\tilde{P}_{|k\rangle_f}$), when the state $|\Phi_0\rangle$ ($|\Phi_1\rangle$) collapses into the state $|k\rangle_f$.  

The following equations are established:
\begin{eqnarray}
  \cos(\theta_{2k+1}-\theta_{2k})&=&\frac{4P_{|2k\rangle_f}-P_{2k}-P_{2k+1}}{2\sqrt{P_{2k}P_{2k+1}}},\nonumber \\
   \sin(\theta_{2k+1}-\theta_{2k})&=&\frac{4P_{|2k+1\rangle_f}-P_{2k}-P_{2k+1}}{2\sqrt{P_{2k}P_{2k+1}}}, \nonumber \\
   \cos(\theta_{2k+2}-\theta_{2k+1})&=&\frac{4\tilde{P}_{|2k\rangle_f}-P_{2k+1}-P_{2k+2}}{2\sqrt{P_{2k+1}P_{2k+2}}},\nonumber \\
   \sin(\theta_{2k+2}-\theta_{2k+1})&=&\frac{4\tilde{P}_{|2k+1\rangle_f}-P_{2k+1}-P_{2k+2}}{ 2\sqrt{P_{2k+1}P_{2k+2}}},\nonumber \\
  \label{theta}
\end{eqnarray}
 The first phase $\theta_0$ is set as zero for the freedom in choosing the global phase. All the phases can be gradually determined.

\textbf{Protocol 2}
--Two new projective measurements onto the orthonormal bases $\mathcal{D}_1$ and $\mathcal{D}_2$ are required to determine the phases uniquely and directly: 

\begin{eqnarray}
\mathcal{D}_{1}&=&\{|2k\rangle_{f1},|2k+1\rangle_{f1},|2k+d\rangle_{f1},|2k+ 1+d\rangle\} \nonumber \\
\mathcal{D}_{2}&=&\{|2k+1\rangle_{f2},|2k+2\rangle_{f2},|2k+ 1+d\rangle_{f2},|2k+ 2+d\rangle\},\nonumber \\
\label{d}
\end{eqnarray}
where $0\le k\le \lfloor\frac{d-2}{2}\rfloor$, and the addition of labels is modulo by $2d$. 

\begin{figure}[ht]

\centering
\includegraphics[scale=0.6]{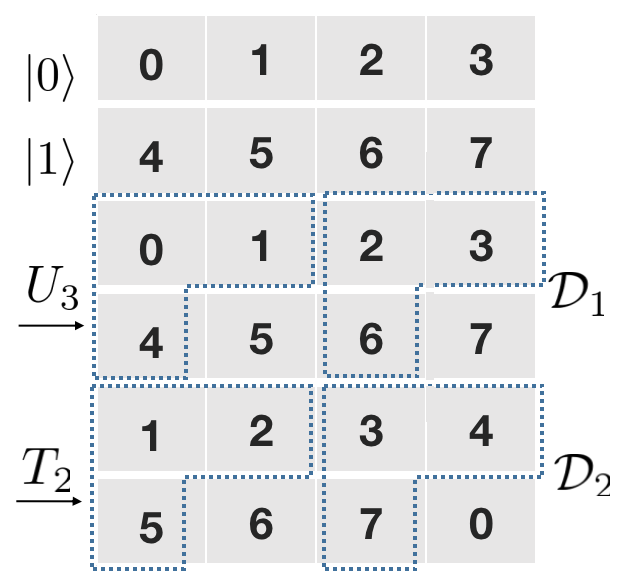}
\caption{Construction of the basis $\mathcal{D}_1$ and $\mathcal{D}_{2}$. 
This picture adopts the four-dimensional case as an example.
Firstly, a partial Fourier transformation on states $\{0,1,4\}$ and $\{2,3,6\}$ is performed to obtain the basis $\mathcal{D}_1$. 
Then a global shift operation $T_2$ is carried out on the basis $\mathcal{D}_1$ to obtain the other basis, $\mathcal{D}_2$. 
$T_2=\sum_{k=0}^{2d-1}|k+1\rangle\langle k|$, where the addition of labels is carried out modulo $2d$. 
}
\label{pic-d12}
\end{figure}

Fig.(\ref{pic-d12}) illustrates the transformation from $\{|0\rangle,|1\rangle\}\otimes \mathcal{B}_0$ into $\mathcal{D}_1$ and $\mathcal{D}_2$.
The subscripts $_{f1}$ and $_{f2}$ denote the FT on the three elements. 
For example, the basis states in $\mathcal{D}_1$ are:

\begin{eqnarray}
|2k\rangle_{f1}     &   =  &  (|2k\rangle+|2k+1\rangle+|2k+d\rangle)/\sqrt{3},  \nonumber    \\
|2k+1\rangle_{f1}   &   =  &  (|2k\rangle+w|2k+1\rangle+w^2|2k+d\rangle)/\sqrt{3},   \nonumber \\
|2k+d\rangle_{f1}   &   =  &  (|2k\rangle+w^2|2k+1\rangle+w|2k+d\rangle)/\sqrt{3}, \nonumber  \\
|2k+ 1+d \rangle      &   =  &  |2k+ 1+d \rangle,
\label{dd1}
\end{eqnarray}
where $w=\exp (i2\pi/3)$.

The product state $|0\rangle\otimes|\phi\rangle$ is measured with the projective measurements on bases $\mathcal{D}_1$ and $\mathcal{D}_2$. 
The probability is recorded as $P_{|k\rangle_{f1}}$ ($P_{|k\rangle_{f2}}$), when the state $|\Phi_0\rangle$  collapses into the $k$-th measurement outcome of each basis. 
With a similar analysis, $\cos(\theta_{k+1}-\theta_k)$  and  $\cos(\theta_{k+1}-\theta_k-2\pi/3)$ can be estimated and thereby, $k=0,\cdots,d-1$. 
The value of $\theta_{2k+1}-\theta_{2k}$ can be determined by $P_{|2k\rangle_{f1}}$ and $P_{|2k+1\rangle_{f1}}$; further, $\theta_{2k+2}-\theta_{2k+1}$ can be determined by 
$P_{|2k\rangle_{f2}}$ and $P_{|2k+1\rangle_{f2}}$. 
 
The above-mentioned protocols need one or two observables to determine the phases uniquely and directly. As far as we know, these protocols use the minimum number of observables among the existing ones. 
Protocol 1 produces at most $5d$ measurement outcomes, similar to  the protocol proposed by Goyeneche et al. \cite{PhysRevLett.115.090401}. 
Protocol 2 needs two settings, however, the number of measurement outcomes is only $80\%$ of the former ones, theoretically. 
As shown in Fig.(\ref{pic-d12}), the unknown pure state $|0\rangle\otimes|\phi\rangle$, in $\mathcal{H}_2\otimes\mathcal{H}_4$ will never collapse into the states $|5\rangle$, $|6\rangle$, and $|7\rangle$, in $\mathcal{H}_8$.
In general, $|\langle \Phi_0 |k\rangle|^2=0$, for $k=d+1, \cdots,2d-1$. 
These projected states $\{|k\rangle,k=d+1,\cdots,2d-1\}$ are simply designed to make the basis complete. 
When experiments are carried out, errors may occur in the quantum state preparations and measurements. 
In this case, if the auxiliary qubit is strictly $|0\rangle$ and the projected states are not completely superposed with states $\{|k\rangle,k=0,\cdots,d-1\}$, the corresponding probabilities will vanish. If not, the measurement outcomes for projected states $\{|k\rangle,k=d+1,\cdots,2d-1\}$ can be observed.

%The first protocol includes  two measurement bases $\mathcal{B}_0$ and $\mathcal{C}_1$, which is the least as far as we know. As for the performance of measurement outcomes, it has the same number as measurements $\{\mathcal{B}_0, \mathcal{B}_1,\mathcal{B}_2,\mathcal{B}_3,\mathcal{B}_4\}$ \cite{PhysRevLett.115.090401}.  To obtain the phase parameters, $\ket{\phi}$ with the ancilla qubit on $\ket{0}$ and $\ket{1}$ are employed, under the $2d$ orthogonal basis states in $\mathcal{C}_1$, it produces 4$d$ outcomes totally.
%The second protocol includes  three measurement basis sets $\mathcal{B}_0$, $\mathcal{D}_1$ and $\mathcal{D}_2$. To obtain the phase parameters, $\mathcal{D}_1$ and $\mathcal{D}_2$ are employed. 
%As shown in Fig. (\ref{pic-d12}), the unknown pure state $|0\rangle\otimes|\phi\rangle$ in $\mathcal{H}_2\otimes\mathcal{H}_4$ would never collapse into the states $|5\rangle$, $|6\rangle$ and $|7\rangle$ in $\mathcal{H}_8$.
%%In general, $|\langle \Phi_0 |k\rangle|^2=0$, for $k=d, d+1, \cdots, 2d-1$. Thus at most  $3d+1$ nonzero probabilities   will appear   in second protocol. Therefore, the second protocol  reduces $d-1$ redundancy and is the least selection as we know.

\emph{Circuit implementation for N-qubit pure state}
--The quantum circuits to implement the bases for both the protocols were designed. 
The projective measurement on a set of orthonormal bases can be translated into a unitary operation with canonical projective measurements onto $\mathcal{B}_0$. 
The operation transforming the bases $\{|0\rangle,\cdots,|j-1\rangle\}$ into $\{|\psi_0\rangle,\cdots,|\psi_{j-1}\rangle\}$ can be denoted as $U$. 
It is established that 
$$\mbox{tr}[\rho|\psi_k\rangle\langle\psi_k|]=\mbox{tr}[\rho(U|k\rangle\langle k|U^{\dag})]=\mbox{tr}[(U^{\dag}\rho U)|k\rangle\langle k|].$$ 
The left side of the expression represents the directly measured probability of the unknown state with basis $\{|\psi_0\rangle,\cdots,|\psi_{j-1}\rangle\}$. 
It is equal to the right side of the expression, obtained by performing the operation $U^{\dag}$, followed by the canonical measurement.  
For $N$-qubits, the projection onto the canonical basis $\mathcal{B}_0$ is implemented by the Pauli $Z$ measurement at each qubit.

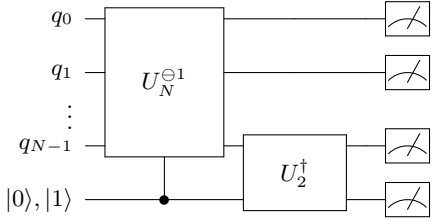
\begin{figure} [!htb]
\[
  \Qcircuit @C=0.8em @R=0.8em {
\lstick{q_0}                                  &\multigate{3}{\quad U_N^{\ominus 1}\quad}  & \qw             &\qw &\meter\\
\lstick{q_1}                                               &\ghost{\quad U_N^{\ominus 1}\quad}    &\qw      &\qw &\meter\\
\lstick{\vdots}                                     &\\
\lstick{q_{N-1}}                                &\ghost{\quad U_N^{\ominus 1}\quad}      &\multigate{1}{\quad U_2^{\dag}\quad}       &\qw &\meter\\
\lstick{|0\rangle,|1\rangle}     &\ctrl{-1} &\ghost{\quad U_2^{\dag}\quad}  &\qw &\meter \\
}
\]
\caption{Quantum circuit for protocol 1. 
This circuit performs a conditional shift operation. 
As shown in Fig.(\ref{U1}), it can be implemented by a single qubit operation $X$ at the auxiliary qubit and a global shift operation, $U_{N+1}^{\ominus1}$. 
}
\label{protocol1}
\end{figure}

\begin{figure} [!htb]
\[
  \Qcircuit @C=0.8em @R=0.8em {
\lstick{q_0}                       &\qw       &\qw &\meter\\
\lstick{q_1}                      &\qw    &\qw &\meter\\
\lstick{\vdots}                                  &\\
\lstick{q_{N-1}}                       &\multigate{1}{ U_3^{\dag}}      &\qw &\meter\\
\lstick{|0\rangle}                       &\ghost{ U_3^{\dag}}       &\qw &\meter\\
}\
\quad\quad\quad
  \Qcircuit @C=0.8em @R=0.8em {
\lstick{q_0}       &\multigate{4}{\quad U_{N+1}^{\ominus 1}\quad} &\qw &\qw &\meter\\
\lstick{q_1}       &\ghost{\quad U_{N+1}^{\ominus 1}\quad}   &\qw   &\qw &\meter\\
\lstick{\vdots}                                  &\\
\lstick{q_{N-1}}      &\ghost{\quad U_{N+1}^{\ominus 1}\quad}  &\multigate{1}{ U_3^{\dag}}   &\qw &\meter\\
\lstick{|0\rangle}    & \ghost{\quad U_{N+1}^{\ominus 1}\quad}     &\ghost{ U_3^{\dag}}             &\qw &\meter \\
}
\]
\caption{Quantum circuit for protocol 2.
}
\label{protocol2}
\end{figure}
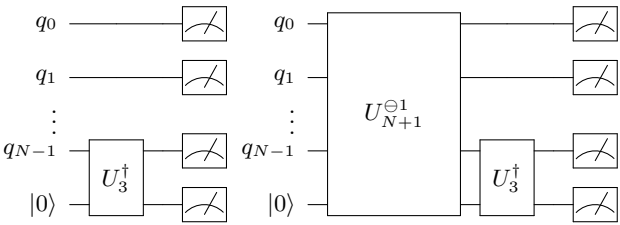

Fig.(\ref{protocol1}) and Fig.(\ref{protocol2}) shows the quantum circuits for protocols 1 and 2, respectively. 
The last line is labeled as auxiliary system and all the measurement settings are on the canonical Pauli $Z$. 

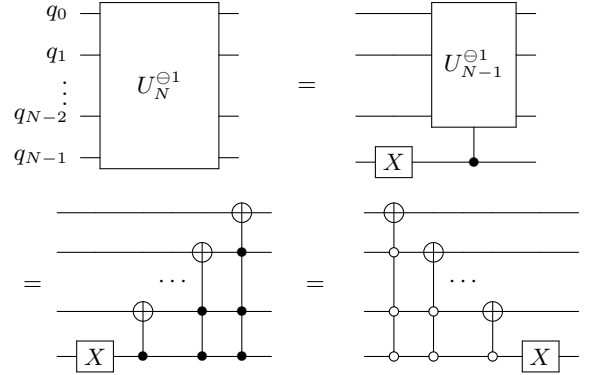
\begin{figure}[!htb]
 \[
 \quad\quad \Qcircuit @C=0.8em @R=0.8em {
\lstick{q_0}  &\multigate{4}{\quad U_N^{\ominus 1}\quad} &\qw&\\
\lstick{q_1}  &\ghost{\quad U_N^{\ominus 1}\quad} &\qw &\\
\lstick{\vdots}&  & & \\
\lstick{q_{N-2}}&\ghost{\quad U_N^{\ominus 1}\quad} &\qw &\\
\lstick{q_{N-1}}&\ghost{\quad U_N^{\ominus 1}\quad} &\qw &\\
}
\quad\quad
\Qcircuit @C=1em @R=1em{
& \quad\quad \\
& \quad\quad\\
& \quad\quad\\
& =\quad\quad\\
& \quad\quad\\
}
\quad
\Qcircuit @C=0.8em @R=0.8em {
&\qw  &\multigate{3}{ U_{N-1}^{\ominus 1}} &\qw&\\
&\qw  &\ghost{U_{N-1}^{\ominus 1}} &\qw &\\
&  & & \\
&\qw&\ghost{ U_{N-1}^{\ominus 1}} &\qw &\\
&\gate{X}&\ctrl{-1} &\qw &\\
}\]
\[
\quad
\Qcircuit @C=1em @R=1em{
& \quad\quad \\
& \quad\quad\\
& \quad\quad\\
& =\quad\quad\\
& \quad\quad\\
}
\Qcircuit @C=0.8em @R=0.8em{
    &\qw      &\qw       &\qw   &   \qw &   \targ      &\qw     \\
   &\qw         &\qw       &\qw & \targ &    \ctrl{-1}  &\qw        \\
 &         &          & \cdots  &        &  \\
&\qw      &\targ       &\qw &  \ctrl{-2} &   \ctrl{-2}  &\qw        \\
    &\gate{X} &\ctrl{-1} & \qw &  \ctrl{-1} &   \ctrl{-1}&\qw &
}
\quad
\Qcircuit @C=1em @R=1em{
& \quad\quad \\
& \quad\quad\\
& \quad\quad \\
& =\quad\quad\\
& \quad\quad\\
}
\quad
\Qcircuit @C=0.8em @R=0.8em {
    &   \targ      & \qw      &  \qw  & \qw      & \qw    & \qw \\
   &    \ctrlo{-1} & \targ    &  \qw  & \qw         & \qw  & \qw \\
  &              &          &   \cdots  &        &  \\
 &   \ctrlo{-2}  &  \ctrlo{-2}& \qw     & \targ     & \qw   & \qw    \\
   &   \ctrlo{-1}  &  \ctrlo{-1} & \qw    &  \ctrlo{-1} &\gate{X}   & \qw
}
\]
\caption{Quantum circuit for implementing $U_N^{\ominus 1}$.
  Two methods of implementation are introduced in this study. The solid (hollow) point denotes that the control qubit is $|1\rangle$ ($|0\rangle$).}
  \label{U1}
\end{figure}

\begin{figure}[!htb]
  \[
 \quad  \Qcircuit @C=0.8em @R=0.8em {
 \lstick{q_{N-1}}                         &\multigate{1}{U_2^{\dag}} &\qw\\
 \lstick{|0\rangle,|1\rangle}   &\ghost{U_2^{\dag}} &\qw 
  }
  \Qcircuit @C=1em @R=1em {
  &\\
  &=\\
  &
  }
  \quad
  \Qcircuit @C=0.8em @R=0.8em {
  &\qw  &\ctrl{1}  &\gate{H}       &\ctrl{1}   &\targ &\ctrl{1} &\qw  \\
    &\gate{H}  &\gate{S^{\dag}}  &\qw   &\targ  & \ctrl{-1} &\targ  &\qw   \\
  }\]
  \[
\Qcircuit @C=0.8em @R=0.8em {
\lstick{q_{N-1}}&\multigate{1}{U_3^{\dag}} &\qw\\
\lstick{|0\rangle}&\ghost{U_3^{\dag}} &\qw
}
\Qcircuit @C=1em @R=1em {
&\\
&=\\
&
}
\quad
\Qcircuit @C=0.8em @R=0.8em {
 &\targ	&\ctrl{1}	 	&\targ	&\ctrlo{1}      &\gate{H}   &\qw     \\
&\ctrl{-1}	&\gate{V_1} 	&\ctrl{-1}	&\gate{V_2} &\ctrlo{-1}  &\qw     \\
}\]
%\Qcircuit @C=0.8em @R=0.8em {
%&\gate{H}&\ctrlo{1}	  &\targ		&\ctrl{1}        &\targ&	\qw\\
%&\ctrlo{-1}&\gate{V_1} &\ctrl{-1}		&\gate{V_2} &\ctrl{-1}&\qw\\
%}\]
  \caption{Quantum circuit for implementing $U_2^{\dag}$ and $U_3^{\dag}$. They are the partial Fourier transformations on two qubits. The one-qubit gates $V_1$ and $V_2$ are mentioned in Eq.(\ref{V12}).}
  \label{U1+2}
\end{figure}
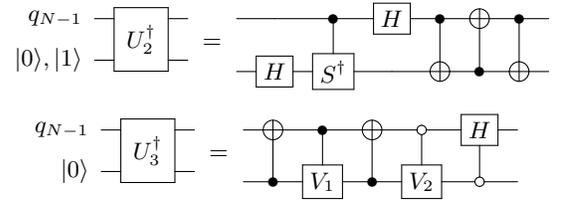

For protocol 1, the basis $\mathcal{C}_1$ is obtained by using partial Fourier transformation and the conditional shift operation, as illustrated in  Fig.(\ref{pic-c1}). 
 $U_N^{\oplus1}$ denotes the quantum version of the increment gate for $N$-qubit. Further, $U_N^{\oplus1}=\sum_{k=0}^{2^N-1}|k+1 \bmod2^N\rangle\langle k|$. Thus, the conditional shift operation $T_1$ is $|0\rangle\langle0|\otimes I+|1\rangle\langle1|\otimes U_N^{\oplus 1}$. 
Let $U_N^{\ominus 1}=(U_N^{\oplus1})^{\dag}$. The circuit implementation is shown in Fig.(\ref{U1}).  
The partial Fourier transformation $U_2$ is a two-qubit operation on the auxiliary qubit and the $N$-th qubit. 
The implementation of its conjugate is shown in Fig.(\ref{U1+2}).

For protocol 2, the basis $\mathcal{D}_1$ is obtained by operating a two-qubit gate $U_3$ in Eq.(\ref{U3matrix}), from the canonical basis. 
\begin{equation}U_3=\frac{1}{\sqrt{3}} \left (\begin{matrix}1 & 1 & 1 & 0   \\1 & w & w^2 & 0 \\1 & w^2 & w & 0 \\0 & 0 & 0 & \sqrt{3}\end{matrix}\right  ).\label{U3matrix}
\end{equation}
The decompositions are expressed as
\begin{equation}
  V_1=\frac{1}{\sqrt{2}}\left (
   \begin{matrix}
       e^{\frac{\pi }{6}i} &  e^{-\frac{\pi }{6}i}   \\
     e^{-\frac{\pi }{3}i}  &  e^{\frac{\pi }{3}i}
     \end{matrix}
    \right  ),
  \\ \\
  V_2=\frac{1}{\sqrt{3}}\left (
   \begin{matrix}
     \sqrt{2} & \sqrt{1}  \\
    \sqrt{1} & -\sqrt{2}
     \end{matrix}
    \right  ).
    \label{V12}
  \end{equation}
%For $U_{N+1}^{\oplus 1}$, the shift operation $T_2$ is applied to the $N$ qubit system. 
The shift operation $T_2$ at $N+1$ qubits system is $U_{N+1}^{\oplus 1}$. 
The basis $\mathcal{D}_1$ is transformed into $\mathcal{D}_2$, through this operation. 

 The quantum circuits for both protocols share a universal form. 
 There are two parts for each protocol. 
 One is related to the partial Fourier transformation between the auxiliary qubit and the $N$-th qubit. 
 The other is related to the quantum version of reduction gate, $U_{N+1}^{\ominus 1}$. 
 As shown in Fig.(\ref{protocol1}), $I^{\otimes N}\otimes |0\rangle\langle0|+U_{N}^{\ominus 1} \otimes |1\rangle\langle 1|=(I^{\otimes N}\otimes X )\cdot U_{N+1}^{\ominus 1}$. 
 The gate $U_{N+1}^{\ominus 1}$ is fundamental in classical computation and is easily realized with one high-dimension photon system \cite{zhao2015experimental}. It is used in the protocol proposed by Goyeneche et al. \cite{PhysRevLett.115.090401}. For a general physical system, it is decomposed into the circuits shown in Fig.(\ref{U1}). Using Lemma 6.1 and 7.2 described in \cite{Barenco1995}, each multi-qubit conditional gate can be decomposed into $O(N^2)$ one-qubit gates and controlled NOT gates.
The tomography circuits for $N+1$-qubit can be obtained from $N$-qubit by performing an additional operation. 
The qubits in the circuits for $N$-qubit pure state tomography are relabeled as $q_1,\cdots,q_{N}$. The new $(N+1)$-th qubit and the auxiliary qubit  are relabeled as $q_0$ and $a$, respectively.
The additional operation is denoted as $X^{q_1\cdots q_{N}a}$, acting on qubit $q_0$. 
If the qubits $q_1,\cdots,q_{N},a$ are all 1, the qubit  $q_0$ will flip.
If not, the qubit $q_0$ will remain.

\emph{Simulation}
--Currently, as quantum devices are still in the NISQ era, noise is inevitably introduced during information processing. To test the feasibility of the proposed protocols, their performances were investigated with a white noise model, which causes decoherence effects. Therefore, the states considered are no longer  pure quantum states.
\begin{eqnarray}
  \rho=(1-\lambda)\ket{\phi}\bra{\phi}+\frac{\lambda}{2^N}I, \label{simu_st}
\end{eqnarray}
where $\lambda$ is the measure of the noise in the device, $I$ is the max-mixture state introduced by the noisy quantum device, and $\ket{\phi}$ is the state to be reconstructed. 

In  the numerical computation performed, $\rho$ is randomly prepared and reconstructed with both protocols 1 and 2. Specifically, there are a total of 100 testing quantum states, drawn from a uniform distribution on Hilbert space. This is implemented by repeatedly applying an operator to $(1-\lambda)\ket{0,\dots,0}\bra{0,\dots,0}+\frac{\lambda}{2^N}I$, where the operator is distributed uniformly according to the Haar measure. The mixed initial state can be realized by driving $\ket{0,\dots,0}$, through a non-unitary channel. 
Then, the fidelities $F$ are calculated by estimating the phases and amplitudes of $\rho_r$ according to protocol 1 and 2, with the size of the system scaling from $1$ to $7$, denoted as $N$.
\begin{eqnarray}
F=\mbox{tr}(\sqrt{\sqrt{\rho_r}\rho\sqrt{\rho_r}})^2.
\end{eqnarray}

\begin{figure}[!htb]
 \includegraphics[width=0.48\textwidth]{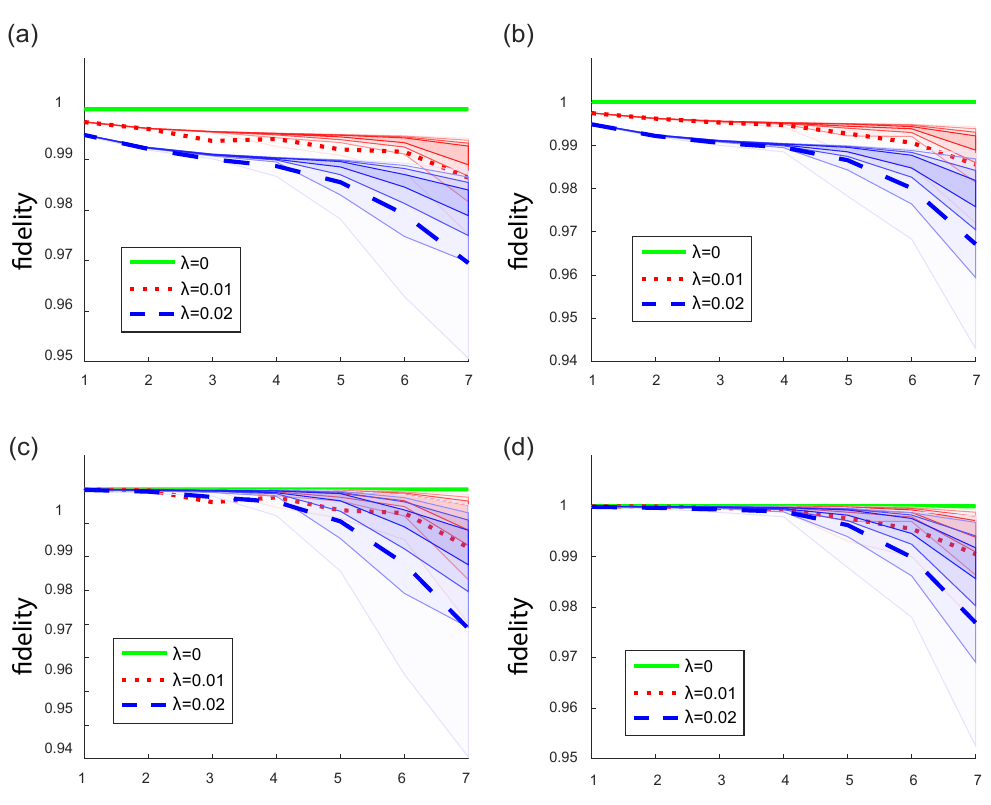}
  \caption{Simulation (Colored online) for protocols 1(a,c) and 2(b,d). (a) and (b) show the fidelity calculated using $\rho$ and its simulation, respectively. (c) and (d) show the fidelity calculated using $\ket{\phi}\bra{\phi}$ and its simulation, respectively.
  The fidelity varies with the $\lambda$ value, the decoherence strength due to the actual quantum device, and the size of the system $N$. The averaged fidelity of 100 random states is shown as a bold line. Percentiles are shown as shaded bands around  the central median lines.}
  \label{simulation1}
\end{figure}

The results are shown in Fig.(\ref{simulation1}). 
In this simulation, different noise levels with $\lambda=0$ (green), $0.01$ (red), and $0.02$ (blue) were studied. 
As indicators of the probability distribution of fidelity, both the averaged and the percentile fidelities were calculated. 
Averaged fidelity, which is the mean of the fidelities of 100 random states, is indicated by bold lines, whereas the dashed, dotted, and solid lines refer to the different noise levels.
Percentiles, shown as shaded bands around the central median lines, are also marked with different color depths. A darker shaded area denotes a value closer to the median points. Thus, percentiles of $i\times 10 \%$($i=1,\dots,9$) are plotted. 
Remarkably, for (a) and (b), $\rho$ is the same as that given in Eq.(\ref{simu_st}); however, for (c) and (d), a decoherence-free simulation is conducted, where $\rho=\ket{\phi}\bra{\phi}$.

The fidelities tend to decay with increasing $\lambda$ and $N$ in both the protocols similarly. 
Moreover, the averaged fidelity and percentiles jointly indicate the shape of the probability distribution of fidelity, as the mean points are always below the median ones. In other words, in most cases, a high-quantity estimation is possible, but the outliers occasionally deviate significantly. We term this distribution as a 'good' probability.
Additional noises will cause a wider probability distribution of the fidelity.
Therefore, if the introduced noise is controllable and limited, both protocols are feasible for producing comparatively good results.

\begin{figure}[!htb]
 \includegraphics[width=0.48\textwidth]{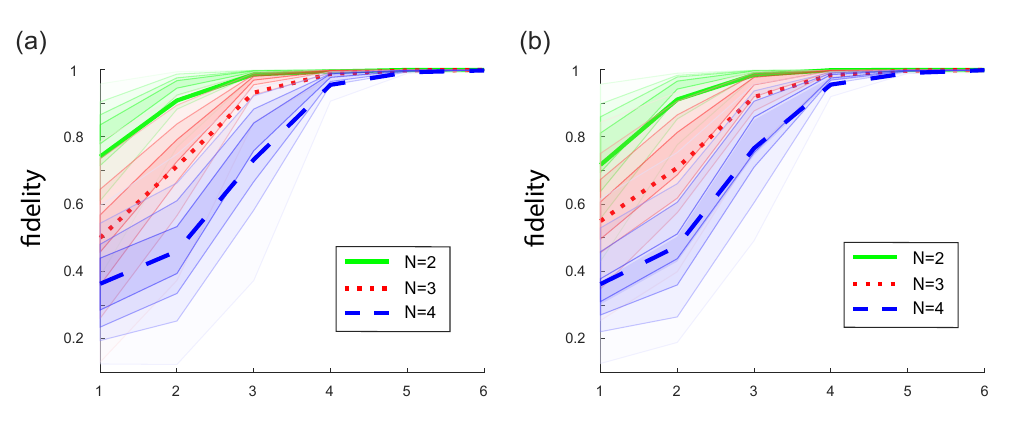}
  \caption{(Colored online) Influence of the number of measurement operations. (a) and (b) are the simulation results for protocols 1 and 2.  When $\lambda=0.003$, the fidelity increases with $M$, the number of measurement operations, denoted as $10^k$($k=1,...,6$). The ideal fidelity is given by $1-\lambda/2^N \approx 1$. The averaged fidelity of 100 random states is indicated as a bold line. Percentiles are shown as shaded bands around the central median lines.}
  \label{simulation2}
\end{figure}

Rigorously speaking, one cannot conduct perfect measurements, as assumed in the aforementioned simulations. Therefore, an additional numerical experiment was performed to evaluate the inaccuracy in estimating the phases and amplitudes with a limited number of measurements. $M$ is the number of projective measurements and takes the value $10^k$ ($k=1,...,6$) in this simulation.  
Therefore, $\lambda$ was set as $0.003$, and the fidelity for 100 initial random states was then calculated, as generated  in the first simulation. The results are shown in Fig.(\ref{simulation2}), with $N= 2$ (green), $3$ (red), and $4$ (blue). The averaged fidelity and percentiles are presented in the same manner as in the first simulation.

The results present a similar consequence as that shown in Fig.~(\ref{simulation1}), where the indicators averaged fidelity and percentiles assist a good probability distribution of fidelity, measured with the proposed protocols. Additionally, when $M$ grows, the results approach $1-\lambda/2^N\approx 1$, which is the assumption for a perfect measurement. Therefore, the inaccuracy in probability will produce little effect on the results, if $M$ takes a sufficiently large value.

\begin{figure}[!htb]
 \includegraphics[width=0.48\textwidth]{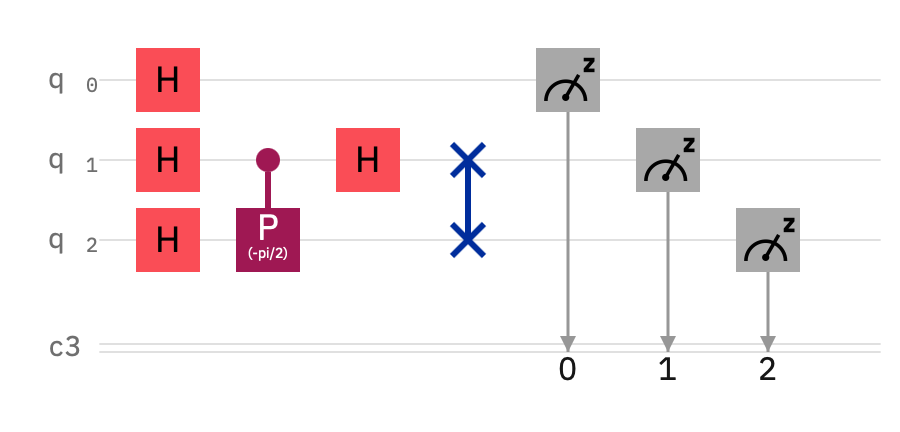}
  \includegraphics[width=0.48\textwidth]{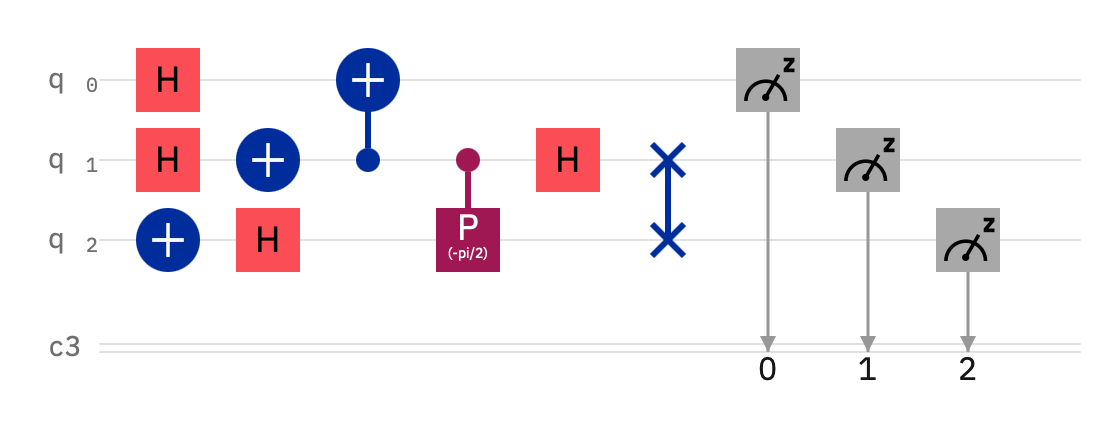}
  \caption{Testing of protocol 1 on ibmq. 
  The two front $H$ gates at qubit 0 and 1 are used to generate the tested pure state. 
  In an actual case, this part is the unknown two-qubit pure state to be determined. 
  The other parts of the circuits are used to determine the phases.
  }
  \label{ibm1}
\end{figure}

\begin{figure}[!htb]
 \includegraphics[width=0.48\textwidth]{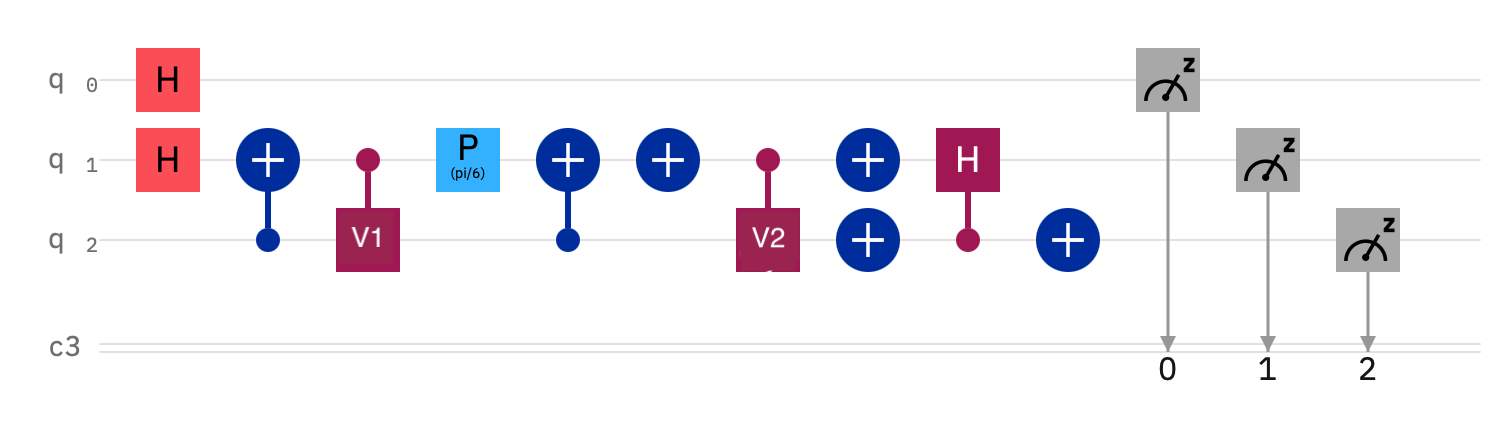}
  \includegraphics[width=0.48\textwidth]{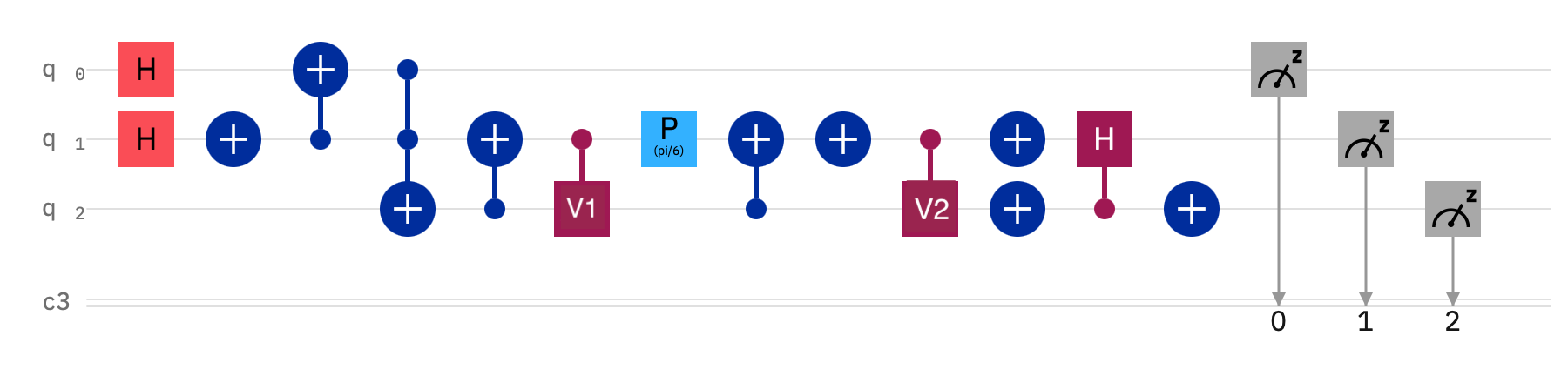}
  \caption{Testing of protocol 2 on ibmq. 
% Also, two $H$ gates at qubit 0 and 1 generate the tested pure state. 
The unitary matrices $V_1$ and $V_2$ are $U(\pi/2,-\pi/2,2\pi/3)$ and $U(2\arccos\sqrt{\frac{2}{3}},0,\pi)$, respectively. 
  The other parts of the circuits are used to determine the phases.
  }
  \label{ibm2}
\end{figure}

\emph{IBMQ simulation}
--The proposed protocols  were tested on  IBM quantum computer ibmq-manila (ibmq)  and ibmq-qasm-simulator (simulator) with 8192
shots. 
We tested the protocols on two-qubit pure state, $(|00\rangle+|01\rangle+|10\rangle+|11\rangle)/2$. 
It is generated by two Hardamard gates on the initial state $|0\rangle^{\otimes2}$.  
The generalized quantum circuits shown in Fig.({\ref{protocol1}}) and Fig.({\ref{protocol2}}) are translated into Fig.(\ref{ibm1}) and Fig.(\ref{ibm2}). 
As for the initial states, each qubit on ibmq is initialized at $|0\rangle$. Protocol 1 measures the initial states $|0\rangle|\phi\rangle$ and  $|1\rangle|\phi\rangle$, while protocol 2 measures the initial states $|0\rangle|\phi\rangle$.

The amplitudes are estimated with the two Pauli Z measurements at each qubit.
The recorded frequencies of the outcomes $0,1,2,3$ are, $\{2089,2005,1984,2114\}$ and $\{2136,1953,2163,1940\}$ for simulator and ibmq, respectively. 
The phases are estimated with the data obtained by the designed circuits, shown in Fig.(\ref{ibm1}) or Fig.(\ref{ibm2}). 
The frequencies are given in Table.(\ref{fibm1}) and Table.(\ref{fibm2}). 
Here we use the value of $\tan(\theta_{k+1}-\theta_{k})$, to estimate the phase difference $\theta_{k+1}-\theta_{k}$. 
For example, from the frequencies of $Z^{\otimes 2}$ and simulator 1 in protocol 1, 
$\tan(\theta_1)=[4\times1028-2089-2005]/[4\times2114-2089-2005]$.  
In protocol 1, the fidelity between the estimated state and target state is $99.98\%$ ($99.65\%$) from the data obtained by the simulator (ibmq). 
In protocol 2, the fidelity $99.89\%$ ($80.13\%$) is obtained from the simulator data (ibmq). 

The fidelity is close to 1 if the data collected by ibmq-qasm-simulator is used. 
This is consistent with the previous numerical experiment. 
The gap between the fidelities of simulator and ibmq can be attributed to the noise when operated on the quantum computer, such as decoherence and depolarization. 
For protocol 2, the fidelity decreases to $80.13\%$  when ibmq is used. 
The main discrepancy appears in the frequencies of the second circuits, as listed in Table.(\ref{fibm2}). 
This is due to the low accuracy of three-qubit Toffoli gate. For the chain structure of the IBM-manila superconducting chip, a comparable deep quantum circuit is required.
This gate can be overlapped if the global quantum increment gate, which is potentially accessible with spatially encoded qubits in the optical system, can be implemented directly \cite{zhao2015experimental}.

\begin{table}
\centering
\caption{Frequency of Protocol 1}
\begin{tabular}{|c|c|c|c|c|}
  \hline
Outcome  		&    0 	&	 1	 & 2 		& 3 		\\
  \hline
  Simulator 1 	& 2114 	& 1028	 & 2056	 & 1014	\\  \hline
  Ibmq 1 		& 2246	& 945	&2022	 &  980	\\ \hline
Simulator 2 	& 2000 	& 998 	& 2035	&1041\\ \hline
  Ibmq 2 		&1953	& 1165	&1763 	&1042  \\
 \hline
\end{tabular}
\label{fibm1}
\end{table}

\begin{table}
\centering
\caption{Frequency of Protocol 2}
\begin{tabular}{|c|c|c|c|c|}
  \hline
Outcome  		&    0 	&	 1	 & 2 		& 3 		\\
  \hline
  Simulator 1 	&  2768	& 696 & 2654	 &697 \\  \hline
  Ibmq 1 		&  2616	& 683 &2488	 & 685	\\ \hline
Simulator 2 	& 2751	& 685 & 681	&700  \\ \hline
  Ibmq 2          & 1867	&654  &891	 & 987 \\
 \hline
\end{tabular}
\label{fibm2}
\end{table}

\emph{Discussion}
--With the development of quantum technologies, efforts are being devoted toward large-scale, fault-tolerant quantum computers. 
It is, therefore, important to design  circuits to uniquely and directly read out the produced quantum states. 
Besides, many applications require the determination of the properties of large-scale complex quantum systems with limited measurement settings, outcomes, and computation resources. 

In this study, we designed two adaptive protocols to determine $N$-qubit pure states. 
The number $N$ denotes the size of the quantum computer, which is arbitrary in the proposed protocols. 
Pure states are the output states of a quantum computer in the absence of noise. If the noise is sufficiently small, the output is close to a pure state.  
On the one hand, traditional Pauli measurements $Z^{\otimes N}$ are used to estimate amplitudes. 
On the other hand, all the phase differences are uniquely and directly calculated with four probabilities obtained via the additional measurement settings for protocols 1 and 2. Theoretically, the number of measurement settings, as well as the number of measurement outcomes, decreased under protocol 2.
Furthermore, quantum circuits for both these protocols were designed, where the quantum version of the reduction gate and the related two-qubit Fourier transformation were required.
Numerical and actual experiments on IBMq indicated the feasibility of both these protocols.

These adaptive protocols can be used for the determination of all unknown $N$-qubit pure states. 
 Notably, we tackled the case where all the amplitudes of $|\phi\rangle$ were nonzero. With randomly chosen  $|\phi\rangle$, there are two ways for the other states to be in a zero measure set. 
The bases to determine the phases can be designed on the projected subspace, or permutation operations can be performed in advance to ensure that the front amplitudes of the new state are nonzero. 
In future research, the design of several fixed and extendable circuits for all quantum states, either pure or mixed, needs to be explored. 
Our protocols are efficient as they provide designs requiring less measurement resources such as settings and outcomes; this is expected to benefit the readout of pure states for future quantum computers.

\textbf{Acknowledgements}
--We thank the IBM Quantum team for making superconducting qubits Manila available via IBM Quantum Experience. 
Y.W. acknowledge the National Natural Science Foundation of China under grant No. 62001260. 
K.L. acknowledge the National Natural Science Foundation of China under grant No. 11905111. K.L. was also supported by the Major Key Project of PCL.

\bibliographystyle{unsrt}
%\bibliography{draft-arxiv}

\begin{thebibliography}{10}

\bibitem{Deutsch1985}
Deutsch, D.  
\newblock {Quantum theory, the Church-Turing principle and the universal quantum computer.} 
\newblock {\em Proc. R. Soc. Lond. A} 400, 97-117 1985.

\bibitem{Deutsch1992}
Deutsch, D. and Jozsa, R. 
\newblock {Rapid solution of problems by quantum computation.} 
\newblock {\em Proc. R. Soc. Lond. A} 439, 553-558 1992.

\bibitem{Shor1994}
Shor, P. 
\newblock {Algorithms for quantum computation: discrete logarithms and factoring.} 
\newblock {\em Proc. 35th Annu. Symp. on Found. of Computer Science} 124-134 (IEEE Comp. Soc. Press, Los Alomitos, CA, 1994

\bibitem{Grover1997}
Grover, L. K. 
\newblock { Quantum computers can search arbitrarily large databases by a single query.} 
\newblock {\em Phys. Rev. Lett.} 79, 4709-4012 1997.

\bibitem{Chuang1998}
Chuang I L, Vandersypen L M K, Zhou X, et al. 
\newblock {Experimental realization of a quantum algorithm.} 
\newblock {\em Nature}.  393(6681): 143-146. 1998.

\bibitem{Ekert1991} 
Ekert A K.
\newblock { Quantum cryptography based on Bell's theorem.}
\newblock {\em Phys. Rev. Lett.}  67(6): 661. 1991. 


\bibitem{Bennett1992} 
Bennett C H, Wiesner S J.
\newblock { Communication via one-and two-particle operators on Einstein-Podolsky-Rosen states.}
\newblock {\em  Phys. Rev. Lett.}  69(20): 2881. 1992.

\bibitem{Bennett1993} 
Bennett C H, Brassard G, Cr{\'e}peau, et al.
\newblock { Teleporting an unknown quantum state via dual classical and Einstein-Podolsky-Rosen channels.} \newblock {\em Phys. Rev. Lett.} 1993, 70(13): 1895.

\bibitem{Gisin2002} 
Gisin N, Ribordy G, Tittel W, et al.
\newblock { Quantum cryptography.} 
\newblock {\em Reviews of modern physics} 74(1): 145. 2002.


\bibitem{Gisin2007} 
Gisin N, Thew R.
\newblock { Quantum communication.}
\newblock {\em Nature photonics}  1(3): 165-171 2007.

\bibitem{Eisert2020}
Eisert J, Hangleiter D, Walk N, et al
\newblock { Quantum certification and benchmarking.} 
\newblock {\em Nature Reviews Physics}  2(7): 382-390. 2020.

\bibitem{Riebe2007}
Riebe M, Chwalla M, Benhelm J, et al. 
\newblock {Quantum teleportation with atoms: quantum process tomography.}
\newblock {\em New Journal of Physics}  9(7): 211. 2007.

\bibitem{Baur2012}
Baur M, Fedorov A, Steffen L, et al. 
\newblock {Benchmarking a quantum teleportation protocol in superconducting circuits using tomography and an entanglement witness.} 
\newblock {\em Phys. Rev. Lett.}  108(4): 040502. 2012.
 
 

\bibitem{Peres2002}
Peres A, Scudo P F, Terno D R. 
\newblock {Quantum entropy and special relativity.} 
\newblock {\em Phys. Rev. Lett.}  88(23): 230402. 2002.

\bibitem{Islam2015}
Islam, R. et al. 
\newblock {Measuring entanglement entropy in a quantum many-body system.} \newblock {\em Nature} 528, 77-83 2015.

\bibitem{Torlai2018}
Torlai G, Mazzola G, Carrasquilla J, et al. 
\newblock {Neural-network quantum state tomography.} 
\newblock {\em Nature Physics}  14(5): 447-450. 2018.

\bibitem{Steffen2006}
Steffen M, Ansmann M, Bialczak R C, et al. 
\newblock {Measurement of the entanglement of two superconducting qubits via state tomography.} 
\newblock {\em Science}  313(5792): 1423-1425.  2006.

\bibitem{Vogel1989}
Vogel K, Risken H. 
\newblock {Determination of quasiprobability distributions in terms of probability distributions for the rotated quadrature phase.} 
\newblock {\em Physical Review A} 40(5): 2847. 1989. 

\bibitem{Leonhardt1995}
Leonhardt U. 
\newblock { Quantum-state tomography and discrete Wigner function.} 
\newblock {\em Phys. Rev. Lett.}  74(21): 4101. 1995.

\bibitem{White1999}
White A G, James D F V, Eberhard P H, et al. 
\newblock {Nonmaximally entangled states: production, characterization, and utilization.} 
\newblock {\em Phys. Rev. Lett.}  83(16): 3103. 1999.

\bibitem{James2005}
James D F V, Kwiat P G, Munro W J, et al. 
\newblock {On the measurement of qubits.}  Asymptotic Theory of Quantum Statistical Inference: Selected Papers. 509-538. 2005. 

  \bibitem{Pauli1933allgemeinen}
  Wolfgang Pauli.
  \newblock {Die allgemeinen prinzipien der wellenmechanik.}
  \newblock In {\em Quantentheorie}, pages 83-272. Springer, 1933.

  \bibitem{Trebino2000}
Trebino R. 
\newblock{Frequency-Resolved Optical Gating: The Measurement of Ultrashort Laser Pulses}
\newblock{Springer} 2000.

\bibitem{peres2006quantum}
  Asher Peres.
  \newblock {\em Quantum theory: concepts and methods}, volume~57.
  \newblock Springer Science \& Business Media, 2006.

  \bibitem{flammia2005minimal}
  Steven~T Flammia, Andrew Silberfarb, and Carlton~M Caves.
  \newblock Minimal informationally complete measurements for pure states.
  \newblock {\em Foundations of Physics}, 35(12):1985-2006, 2005
  
  
  \bibitem{Vogt1978}
Vogt A. 
\newblock {Position and momentum distributions do not determine the quantum mechanical state.}
\newblock {\em Mathematical Foundations of Quantum Theory} Academic Press,  365-372. 1978.

\bibitem{Moroz1983}
Moroz B Z. 
\newblock {Reflections on quantum logic.} 
\newblock {\em International Journal of Theoretical Physics} 22(4): 329-340. 1983.

\bibitem{Moroz1994}
Moroz B Z, Perelomov A M. 
\newblock {On a problem posed by Pauli.} 
Theoretical and Mathematical Physics, 1994, 101(1): 1200-1204.

\bibitem{Corbett2006}
Corbett J V. 
\newblock {The Pauli problem, state reconstruction and quantum-real numbers.}
\newblock {\em Reports on Mathematical Physics} 57(1): 53-68. 2006.

\bibitem{Heinosaari2013}
Heinosaari T, Mazzarella L, Wolf M M. 
\newblock {Quantum tomography under prior information.} 
\newblock {\em Communications in Mathematical Physics} 318(2): 355-374. 2013
   

  \bibitem{Jaming2014uniqueness}
  Philippe Jaming.
  \newblock Uniqueness results in an extension of Pauli's phase retrieval
    problem.
  \newblock {\em Applied and Computational Harmonic Analysis}, 37(3):413-441, 2014. 
  
  \bibitem{Carmeli2016}
  Carmeli C, Heinosaari T, Kech M, Schultz J and  Toigo A.
  \newblock Stable pure state quantum tomography from five orthonormal bases.  
  \newblock {\em Europhysics Letters} 115.3 30001, 2016.



  \bibitem{Oren2017}
Oren D, Mutzafi M, Eldar Y C, et al. 
 \newblock Quantum state tomography with a single measurement setup. 
 \newblock {\em Optica},  4(8): 993-999 2017.

\bibitem{Song2017}
Song C, Xu K, Liu W, et al. 
\newblock 10-qubit entanglement and parallel logic operations with a superconducting circuit. 
\newblock {\em Phys. Rev. Lett.}, 119(18): 180511 2017.


\bibitem{Lundeen2011}
Lundeen J S, Sutherland B, Patel A, et al. 
  \newblock Direct measurement of the quantum wavefunction. 
  \newblock {\em Nature},  474(7350): 188-191 2011.

\bibitem{Lundeen2012}
 Lundeen J S, Bamber C. 
\newblock Procedure for direct measurement of general quantum states using weak measurement. 
\newblock {\em Phys. Rev. Lett.}, 108(7): 070402 2012.

\bibitem{Thekkadath2016}
Thekkadath G S, Giner L, Chalich Y, et al. 
\newblock Direct measurement of the density matrix of a quantum system. 
\newblock {\em Phys. Rev. Lett.}, 117(12): 120401 2016.

\bibitem{Calderaro2018}
Calderaro L, Foletto G, Dequal D, et al. 
\newblock Direct reconstruction of the quantum density matrix by strong measurements. 
\newblock {\em Phys. Rev. Lett.},  121(23): 230501 2018. 

\bibitem{Zhang2019}
Zhang S, Zhou Y, Mei Y, et al. 
\newblock $\delta$-quench measurement of a pure quantum-state wave function. 
\newblock {\em Phys. Rev. Lett.},  123(19): 190402 2019.





\bibitem{Chen2013}
Chen J, Dawkins H, Ji Z, et al. 
\newblock Uniqueness of quantum states compatible with given measurement results.
\newblock {\em Physical Review A},  88(1): 012109 2013.
 
  \bibitem{Finkelstein2004}
  Finkelstein J. 
  \newblock Pure-state informationally complete and 'really' complete measurements. 
  \newblock {\em Physical Review A}, 70(5): 052107 2004.
  
    \bibitem{wang2018pure}
  Yu~Wang and Yun Shang.
  \newblock Pure state really informationally complete with rank-1 povm.
  \newblock {\em Quantum Information Processing}, 17(3):51, 2018.
  
    \bibitem{PhysRevLett.115.090401}
  D.~Goyeneche, G.~Ca\~nas, S.~Etcheverry, E.~S. G\'omez, G.~B. Xavier, G.~Lima,
    and A.~Delgado.
  \newblock Five measurement bases determine pure quantum states on any
    dimension.
  \newblock {\em Phys. Rev. Lett.}, 115:090401, 2015.  
  
  
    \bibitem{Zambrano2020}
  Zambrano L, Pereira L, Mart{\'\i}nez D, et al. \newblock Estimation of pure states using three measurement bases. 
   \newblock {\em Physical Review Applied}, 14(6): 064004 2020.


  
   \bibitem{Harrow2009}
 \newblock Harrow A W, Hassidim A, Lloyd S. Quantum algorithm for linear systems of equations. 
 \newblock {\em Phys. Rev. Lett.}, 103(15): 150502 2009.
  

  
   %\bibitem{stefano2017determination}
 % Quimey~Pears Stefano, Lorena Reb{\'o}n, Silvia Ledesma, and Claudio Iemmi.
  %\newblock Determination of any pure spatial qudits from a minimum number of measurements by phase-stepping interferometry.
  %\newblock {\em Physical Review A}, 96(6):062328, 2017.

  %\bibitem{stefano2019set}
  %Quimey~Pears Stefano, Lorena Reb{\'o}n, Silvia Ledesma, and Claudio Iemmi.
  %\newblock Set of 4d--3 observables to determine any pure qudit state.
  %\newblock {\em Optics letters}, 44(10):2558--2561, 2019.

 
  
  
   % \bibitem{zambrano2019improved}
 % L~Zambrano, L~Pereira, and A~Delgado.\newblock Improved estimation accuracy of the 5-bases-based tomographic method.\newblock {\em Physical Review A}, 100(2):022340, 2019.

%\bibitem{Caves2002} Caves C M, Fuchs C A, Schack R.  \newblock {Unknown quantum states: the quantum de Finetti representation.} \newblock {\em Journal of Mathematical Physics} 43(9): 4537-4559. 2002. 

% \bibitem{Renes2004} Renes J M, Blume-Kohout R, Scott A J, et al.  \newblock {Symmetric informationally complete quantum measurements.}  \newblock {\em Journal of Mathematical Physics} 45(6): 2171-2180.  2004.
 

%\bibitem{Geng2021}
%Geng I J, Golubeva K, Gour G. 
%\newblock {What Are the Minimal Conditions Required to Define a Symmetric Informationally Complete Generalized Measurement?} 
%\newblock {\em Physical Review Letters}  126(10): 100401. 2021.

  %\bibitem{vrehavcek2004minimal} Jaroslav {\v{R}}eh{\'a}{\v{c}}ek, Berthold-Georg Englert, and Dagomir Kaszlikowski. \newblock Minimal qubit tomography. \newblock {\em Physical Review A}, 70(5):052321, 2004.

  

  \bibitem{zhao2015experimental}
  Yuan-yuan Zhao, Neng-kun Yu, Pawe{\l} Kurzy{\'n}ski, Guo-yong Xiang, Chuan-Feng
    Li, and Guang-Can Guo.
  \newblock Experimental realization of generalized qubit measurements based on
    quantum walks.
  \newblock {\em Physical Review A}, 91(4):042101, 2015.
  
   \bibitem{Barenco1995}
  Barenco A, Bennett C H, Cleve R, et al. 
  \newblock Elementary gates for quantum computation.  \newblock {\em Physical review A}, 52(5): 3457 1995.

  \end{thebibliography}

\end{document}